\documentclass[useAMS,usenatbib,fleqn]{mn2e}
\usepackage{amsmath,amssymb,amsfonts,latexsym}
\usepackage[dvips]{graphicx}
\usepackage{longtable}
\usepackage{fixltx2e}
\usepackage{natbib}
\usepackage{hyperref}

\newcommand{\be}{\begin{equation}}
\newcommand{\ee}{\end{equation}}

\def\be{\begin{equation}} 
\def\ee{\end{equation}} 
\def\beq{\begin{eqnarray}} 
\def\eeq{\end{eqnarray}}

\def\04a{{2004 a}}
\def\04b{{2004 b}}

\title{Magneto - Thermal Evolution of Neutron Stars with Emphasis to Radio Pulsars}

\author[U.~Geppert] {U.~Geppert\thanks{E-mail:ulrich.geppert@dlr.de}\\
J. Gil Institute of Astronomy, University of Zielona G\'{o}ra, ul. Szafrana 2, 65-516, Zielona G\'{o}ra, Poland }

\begin{document}

\date{}

\maketitle

\label{firstpage}

\begin{abstract}
The magnetic and thermal evolution of neutron stars is a very complex process with many, also non-linear, interactions. For a decent understanding of neutron star physics, these evolutions cannot be considered isolated.  A brief overview is presented, which describes the main magneto - thermal interactions that determine the fate both of isolated neutron stars and accreting ones.\\ 
\noindent Special attention is devoted to the interplay of thermal and magnetic evolution at the polar cap of radio pulsars. There, a strong meridional temperature gradient is maintained over the lifetime of radio pulsars. It may be strong enough to drive thermoelectric magnetic field creation which perpetuate a toroidal magnetic field around the polar cap rim. Such a local field component may amplify and curve the poloidal surface field at the cap,  forming a strong and small scale magnetic field there as required for the radio emission of pulsars.
 \end{abstract}

\begin{keywords}
stars: neutron - stars: magnetic fields - pulsars: general - stars: interiors
\end{keywords}

\section{Preface}
When I started to try to understand some aspects of neutron star physics, one of the first papers I read was \cite{SvdH82} about the evolution of millisecond pulsars. Then the connection between the (at that time in optical light)  invisible pulsars and the fascinating manifestations of supernova remnants \citep{SDB84} increased my interest in this branch of astrophysics.  The observation of millisecond pulsars in the early eighties was a challenge, which was met by Prof. Srinivasan and his colleagues (see e.g. \cite{BS86,S89}). Later the pioneering work of \cite{SriniBMT90} triggered my interest to understand how the very long-lived core magnetic field is affected by the rotational history of neutron stars and how the evolutions of core and crustal field are coupled to each other. Meanwhile Prof. Srinivasan enriched our knowledge about neutron stars by many contributions, among others about their evolution in accreting binary systems, about the recycling of old pulsars to millisecond pulsars, about the supernova remnants and pulsar wind nebulae which are results of neutron star births, about the use of radiotelescope for pulsar observations  and about many other topics. As far as I can see being a far distant observer, Prof. Srinivasan succeeded also in the formation of a group of excellent and worldwide renown Indian scientists working in the field of neutron star physics.
May be there exists a not yet discovered age-relativistic effect that makes the time running faster when seen by an aging person. It seems to be this effect that made me surprised to recognize that Prof. Srinivasan celebrates already his 75th birthday. This special issue is a good opportunity to esteem his life-work and I am happy that I can contribute to it.
 
\section{The many facets of magneto - thermal interactions}
There is certainly a general consensus that in order to get a realistic idea about neutron star physics, their magnetic and thermal evolution  can not be considered isolated. The magneto - thermal evolution of neutron stars is a multifaceted process. It starts already at a neutron star's birth during the proto - neutron star phase, when powerful convective motions may drive a dynamo - like amplification of magnetic fields up to magnetar field strengths exceeding $10^{16}$ G in the core and inner crust region. Convection will be driven by strong lepton and temperature gradients which together with sufficiently fast and differential rotation of the proto - neutron star enables via $\alpha - \Omega$ dynamos very efficient magnetic field growth \citep{TD93,MPU02}.\\

\noindent In magnetars, which are young (age $\lesssim 10^5$ yrs) and highly magnetized neutron stars, such a superstrong internal magnetic field can power the bursting activities by building up enormous magnetic stresses in the crystallized crust \citep{PerP11,L15,GKLH15}. It provides also by its Ohmic decay in the crust the Joule heat that explains their high X-ray luminosity. Outburst activities observed at magnetars may also be caused by twisting and subsequent rapid reconnection of magnetospheric fields \citep{TLK02,PBH12}. Magnetospheric processes, however, are only indirectly coupled to the thermal evolution of neutron stars. This coupling may proceed via the temperature dependence of the electric conductivity which, in turn, controls the magnetic field decay in the crust.  The thermal evolution of magnetars is almost completely controlled by the amount of magnetic energy stored in the crust. In this about $1$ km thick layer is the electric resistivity large enough to guarantee ample transfer of magnetic into thermal energy which holds the surface temperature well above $T_s \sim 10^6$ K. \citep{PLMG07, VRPPAM13}.\\

\noindent There is also another effect, where a strong crustal magnetic field influences the thermal evolution. The heat flux perpendicular to the magnetic field is suppressed in comparison to the heat flux parallel to the magnetic field by $(\omega_B\tau)^2$, where $\omega_B$ and $\tau$ are the electron gyrofrequency and relaxation time, respectively. This magnetization parameter can in magnetar crusts easily exceed $\omega_B\tau\gtrsim 1000$ (see e.g. \cite{GR02}), i.e. the perpendicular heat flux is impeded drastically. Thus, either a strong toroidal field in the deeper crustal layers and/or the poloidal component of the crustal magnetic field in a quite big region north and south of the magnetic equator cause a significant slower cooling than in the non-magnetic case \citep{PGK07,APM08a,APM08b,PVPR13}.\\

\noindent During the first $\sim 10^5$ years of their life, neutron stars are predominantly cooled by neutrino emission. Since the neutrino emissivity is affected by the internal magnetic field, there appears another facet of the magneto - thermal interaction. With increasing local core magnetic field the neutrino loss rate increases too. Naturally, the strongest effect of the magnetic field on neutrino emissivities can be expected where the field is strongest - in magnetars \citep{RRED05,PCY07}. However, since the magnetic field varies its strength within the core it is unclear whether a significant net enhancement of the neutrino emissivity can be expected. At the very beginning of a neutron star's life neutrino-nucleon scattering and electron neutrino emission/absorbtion in the dense and highly magnetized medium determine the opacity for neutrino emission and therefore the early cooling process \citep{AL99}. Core matter consists of neutrons, protons, electrons and perhaps of an admixture of ``exotic'' particles.  The proton fraction determines whether the cooling proceeds very fast via direct URCA processes or slower via modified URCA processes. The separation between both processes is a proton fraction of $11\%$; larger fractions will be present in neutron stars whose mass exceeds $1.4$M$_{\odot}$ (see \cite{PGW06}). Recently, \cite{CCN16} studied the influence of a superstrong ($>10^{18}$G) core magnetic field on to the proton fraction. Such a strong magnetic field will modify the phase space. This results in a sufficiently large proton fraction that allows direct URCA processes also in neutron stars with masses below $1.4$M$_{\odot}$. In this way a very strong internal magnetic field could accelerate the early cooling of ``low mass'' neutron stars. The neutrino emissivity is sensitive to the appearance of a phase transition from normal into superconductive/superfluid one \citep{KYH97}. These phase transitions, characterized by pairing of Fermions (protons, neutrons), affect seriously the thermal evolution of neutron stars. Since during the pair formation neutrinos are emitted, the cooling proceeds by pair breaking and formation \citep{PGW06} faster than in regimes where no pairing can occur. Some pairing states are energetically favored in strong ($10^{16...17}$ G) fields \citep{SWW98}. On the other hand, if the neutron star has been cooled down so much that big parts of the core and inner crust are well in a superfluid/superconductive state, the neutrino emissivity will be considerably suppressed \citep{PGW06}. Once, after the transition from normal to superfluid matter, the core magnetic field is organized in flux tubes and the neutrons in vortices, the core magnetic field evolution proceeds in a much more complicated way \citep{BPP69,EP79}. The process of magnetic flux expulsion from the core is then affected by interaction of the flux tubes with neutron vortices (whose number increases with the rotational frequency of the star), scattering off normal electrons, buoyancy of the tubes, and their own tension \citep{SriniBMT90,DCC93, KG01,GAGL15,EPRGV16}. All these processes are still subject of intense scientific debates.\\

\noindent Ambipolar diffusion of magnetic fields is the joint transport of magnetic flux and charged particles (protons and electrons) relative to neutral background particles (neutrons). This process has been studied for the first time in the context of neutron star magnetic field decay by \cite{GR92}. It is a dissipative process that may dominate the evolution of the magnetic field maintained by currents in the core and is another process which is tightly connected to the thermal evolution  of neutron stars \citep{HRV08,HRV10,GJS11}. In the pioneering work of \cite{GR92}, the background neutrons in the core were assumed to be immobile with respect to the with the field co-moving charged particles. The ambipolar drift velocity can be divided into a solenoidal and an irrotational component, where the time scale of the former decreases while the time scale of the latter increases  in the process of core cooling. Very recently \cite{GKO17} presented a study that relaxes the assumption of immobile neutrons. In their analysis the regime of irrotational ambipolar diffusion does not appear. The thermal state of the normal core matter affects the time scales of ambipolar diffusion via the temperature dependencies of the electron-neutron and proton-neutron relaxation times, the periods of time between two collisions of the respective particles. After the phase transition of the core matter into a superfluid/superconducting state, neutrons and protons are preferentially organized in vortices and fluxtubes. Then, the characteristic time scale of magnetic field decay by ambipolar diffusion is not longer determined by frictional forces between normal protons, neutrons, and electrons but by the mutual friction of vortices and fluxtubes.  When and where core matter undergoes a transitions from the normal into the superfluid/superconducting  state determines crucially the influence of ambipolar diffusion on the global magnetic field evolution. If both fast neutrino reactions (as appear in massive cores) and an early transition into superfluid phases (already at temperatures $\lesssim 10^9$ K) occur, this dissipative process may cause especially in magnetars advection that substantially reorganizes their internal field structures \citep{PAPM17}.\\

\noindent A wide field of observations that can only be explained by intense magneto - thermal interaction came up when  anisotropies of the surface temperature of isolated neutron stars were found. By use of the analogy with sun spots \cite{FM72} discussed the presence of hot spot at pulsar surfaces as a magnetic field effect.  First evidences provided X-ray observations obtained by the EXOSAT satellite in the eighties. They triggered the pioneering work of \cite{GH83}, who interpreted the modulations seen in the light curve as being caused by  the misalignment of rotation and dipolar magnetic field axis. More and more reliable evidences for a two-component blackbody spectrum were found in ROSAT data \citep{O95}. The spectral fits for at least three relatively young neutron stars revealed that the emitting area of the hotter component appears to be significantly smaller than the whole neutron star surface. Theoretical explanations rely on the fact that the heat conductivity becomes a tensor in the presence of magnetic fields. Thus, the heat flux parallel to the magnetic field can flow practically unimpeded while perpendicular to the field the heat flux is strongly suppressed. As already mentioned, this suppression which is proportional to the square of the magnetization parameter, can reduce the perpendicular heat flux easily by a factor of $10^6$ in the neutron star crust. Various authors studied this phenomenon \citep{PSVZ94, ZPSV95,GKP04,GKP06,PMP06,PVPR13}. A particularly impressive example for the intensive magneto - thermal interaction is the creation of small hot spots at the surface of neutron stars as consequence of the Hall drift appearing in the crust of highly magnetized neutron stars \citep{VRPPAM13,GV14}. It forces the creation of magnetic spots, characterized by a strong but small scale surface field. The crustal region beneath these spots is a site of intense Joule heating. Since within and below the spot the field is predominantly perpendicular oriented to the surface, this heat outflows easily, making the magnetic spot also a hot one. In the surrounding of the spot, however, the field lines within the crust are strongly curved into parallel direction with respect to the surface, thereby acting as a barrier to the radial heat flux. Thus, the hot spot is encircled by a significantly cooler surface region. By creating out of large scale magnetic field structures significantly shorter ones in the crust, the Hall drift may cause pronounced anisotropies in the surface temperature distribution.\\

\noindent Magneto-thermal processes play a crucial role for accreting neutron stars. In binary systems formed with main sequence stars or white dwarfs, accretion onto the surface of neutron stars may dramatically change their magneto - thermal and rotational evolution. While a weakly magnetized neutron star will accept accreting matter on large parts of its surface, the presence of a strong dipolar field mediate the accretion process by channeling the inflowing matter towards the magnetic poles. The transfer of angular momentum onto or from the neutron star determines its spin-up or spin-down \citep{GPL77}. The spin-up of old pulsars to millisecond pulsars by heavy accretion during the spiral-in period in binary systems has been first studied   by \cite{SvdH82} (see also \cite{BvdH91}). The relation to the thermal evolution is given by the possibly strong heating of crustal layers by the accreted material \citep{R90,C92,GU94,KB97,UGK98,KB99,IHR08,BD13}. Heating of the crust proceeds not only by compression of crustal matter, exerted by the burden of the accreted material. The main heat sources are pycnonuclear reactions. They change the elements according to the density which toward the core sinking accreted material reaches \citep{HZ90,UR01,WDP13}. An increase of the crustal temperature by the assimilation of accreted matter into the crustal structure reduces the electric conductivity there. Thus, the currents that maintain the crustal magnetic field decay faster which in turn affects the accretion process by the weakening of the external field that channels the inflowing matter.\\

\noindent Given the existence of strong temperature gradients in certain layers and during certain periods of neutron star life, the transfer of thermal energy into magnetic is a genuine process of the magneto - thermal evolution scenarios. The transfer of thermal into magnetic energy may proceed either via an instability \citep{BAH83,ULY86,GW91}, or the temperature gradient is sustained by ``external'' processes as e.g. the bombardment of the polar cap of radio pulsars. Then the thermoelectrically generated electric field acts as a battery. The potency of thermoelectric effects onto the magneto - thermal evolution of neutron stars relies on the fact that electric charges are carriers of both thermal and electric currents, it depends crucially on the maintenance of sufficiently strong temperature gradients. Enormous temperature gradients are created at the very beginning of neutron star life during the thermal relaxation phase when the core cools faster by neutrino emission than the crust \citep{GYP01}. However, the huge neutrino emissivity during the first year after neutron star birth prevents the emergence of a temperature gradient perpendicular to the strong radial one which is necessary for the instability to operate (Vigan\`{o}, D. \& Pons, J.A. 2012, private communication). Another site where strong temperature gradients are maintained over the first $\sim 1000$ years of neutron star life is the outer crust, sometimes named the envelope, at densities $\lesssim 10^8$ g cm$^{-3}$. At so early times this layer is in a liquid state. Solving the coupled induction and heat transfer equation, an instable growth of a small scale ($\sim 100$ m) toroidal field component on a time scale of $\sim 100$ days could be found \citep{BAH83,ULY86,WG92}, as well as a rapid growth of large scale poloidal components \citep{GW95}. However, a simulation of a growth of the poloidal dipolar field to strengths $\gtrsim 10^{12}$ G could not be provided. For fields of this strength a coupling of the induction equation not only to the heat transfer equation but also to the equations of hydrodynamics is necessary because then the Lorentz forces are strong enough to drive circulations in the liquid shell. This numerical challenge has not yet been accepted by anyone. Thermoelectric interactions have not received much attention in recent years. Obviously, the conviction has prevailed that they can not play an important role in the life of a neutron star.\\

\noindent As the probably incomplete brief description of the facets of the magneto - thermal evolution of neutron stars proves, it needs rather a textbook than a short overview to describe all of them adequately and to discuss our present understanding. Therefore I want to discuss here a possible appearance of the almost forgotten thermoelectric effects at the polar cap of radio pulsars. A special volume to honor Prof. Srinivasan should be a good place to present an idea which is not yet completely elaborated and has still some speculative elements.\\

\noindent In the following, the intense magneto - thermal interactions at the polar cap of radio pulsars will be discussed in the context of the partially screened gap (PSG) model, an extension of the vacuum gap model for radio pulsars proposed by \cite{RS75}. The appearance of a strong and enduring meridional temperature gradient makes it probable that thermoelectric effects may influence the local magnetic field structure at the very surface of the cap. By an estimate of the leading terms of the induction equation, the Ohmic decay and the thermoelectric battery term, the feasibility of a significant thermoelectric field effect at the polar cap surface is assessed.

\begin{figure*}
\centering
\includegraphics[width=8cm, height=5cm]{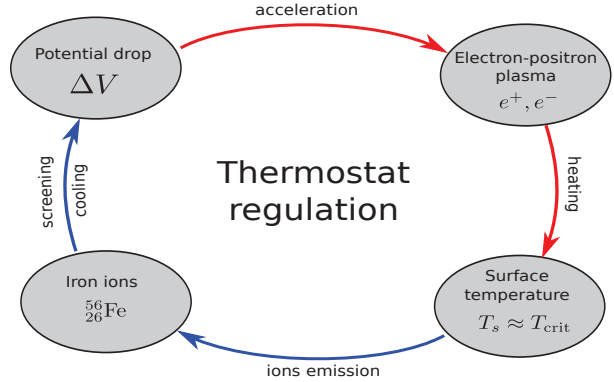}
\caption{Schematic presentation of the thermostatic regulation at the polar cap. Figure taken from \citealt{SMG15}.}
\label{fig:thermostat}
\end{figure*} 
\section{Partially screened gap}
\noindent The vacuum gap model of radio pulsar emission \citep{RS75} is based on the formation of an electric potential gap above the polar cap within which the charges are accelerated, which eventually emit the observed radio waves. Basic ingredient of this models is a sufficiently high cohesive energy that may retain charges in the surface layer of the cap to allow the emergence of a potential gap. The vacuum gap model predicts the $\vec{E}\times \vec{B}$ drift of the electron-positron plasma around the local magnetic pole. This carousel of spark plasma filaments can be observed as so-called drifting subpulses  in the radio lightcurves of many pulsars. However, the predictions of the vacuum gap model yielded much faster drift rates of subpulses than were found in the observations  \citep{vLSRR03,GS03}. In order to reconcile observations with theoretical predictions the partially screened gap model (PSG) has been developed.\\
\noindent Its basic idea is that the thermostatically regulated surface temperature at the polar cap surface allows during the formation phase of the potential gap an ``additional'' thermionic outflow of ions, which screen the gap and reduce in that way the drift velocity of the sparks (see Fig.~\ref{fig:thermostat}). This explains well the phenomenon of drifting subpulses. 
\noindent Two components of the neutron star magnetic field determine the rotational history  and the ability to emit radio waves of pulsars. Its global far-reaching dipolar field $B_d$, estimated from the rotational period $P$ and its time-derivative $\dot{P}$, rules the gradual braking of the pulsar rotation. The local small scale field at the very surface of the polar cap $B_s$ has to exceed $\sim 5\cdot 10^{13}$G in order to guarantee a sufficiently high cohesive binding of charges in the surface of the polar cap. As can be seen in Tab.~\ref{tab:xray_all}, simultaneous radio and X-ray observations support the PSG; the flux conservation argument returns for the small scale magnetic field at the hot plar cap surface $B_s\sim (10\ldots 100)\times B_d$, as required for a sufficiently yielding pair production. \\

\begin{table*}
    \caption{Observed spectral properties of rotation-powered normal pulsars with polar cap X-ray emission. 
     }
    \label{tab:xray_all}
    {\footnotesize
    \begin{center}
   \begin{tabular}{llcccccccc}
        \hline
        \hline
        No. &  Name    &   $P$   &   $\dot{P}$   &   $B_{\rm d}$   &  $R_{dip}$  &  $R_{\rm pc}$   &    $T_{\rm s}$   &  $ B_{\rm s}$ & Ref.  \\
        &  &   {$\left ( {\rm s} \right )$}   &   {$\left ( 10^{-15}\,{\rm s\,s^{-1}} \right )$}   &   {$\left ( 10^{12}\,{\rm G} \right )$}   & {$\left ( {\rm m} \right )$}  &  {$\left ( {\rm m} \right )$} &  {$\left ( {\rm 10^{6} K} \right )$} &  {$\left ( 10^{14}\,{\rm G} \right )$}  \\
        \hline
 & & & & & & & & &  \\
1  &  J0108--1431   &   0.808   &    0.077   &    0.504   &  $161$    &  $33^{+24}_{-14}$ &  $1.7^{+0.3}_{-0.1}$   &   $0.12^{+0.24}_{-0.08}$   &   1 \\
2  &  B0355+54   &   0.156   &    4.397   &    1.675   &  $366$  &  $92^{+123}_{-54}$ &  $3.0^{+1.5}_{-1.1}$   &   $0.27^{+1.27}_{-0.22}$   &   2 \\
3  &  B0628--28   &   1.244   &    7.123   &    6.014   &  $130$  &  $64^{+70}_{-50}$ &  $3.3^{+1.3}_{-0.6}$   &   $0.25^{+4.88}_{-0.19}$   &   3 \\
4  &  J0633+1746   &   0.237   &    10.971   &    3.258   &  $297$  &  $36^{+9}_{-9}$ &  $2.3^{+0.1}_{-0.1}$   &   $2.21^{+1.83}_{-0.82}$   &   4 \\
5  &  B0834+06   &   1.274   &    6.799   &    5.945   &  $128$  &  $30^{+56}_{-15}$ &  $2.0^{+0.8}_{-0.6}$   &   $1.05^{+3.19}_{-0.92}$   &   5 \\
6  &  B0943+10   &   1.098   &    3.493   &    3.956   &  $138$  &  $20^{+4}_{-4}$ &  $3.1^{+0.3}_{-0.2}$   &   $1.99^{+0.96}_{-0.62}$   &   6 \\
7  &  B1133+16   &   1.188   &    3.734   &    4.254   &  $133$  &  $14^{+7}_{-5}$ &  $2.9^{+0.6}_{-0.4}$   &   $3.9^{+1.12}_{-0.76}$   &   7 \\
8  &  B1451--68   &   0.263   &    0.098   &    0.325   &  $282$  &  $14^{+24}_{-12}$ &  $4.1^{+1.4}_{-0.8}$   &   $1.36^{+113.57}_{-1.18}$   &   8 \\
9  &  B1719--37   &   0.236   &    10.854   &    3.234   &  $298$  &  $237^{+391}_{-123}$ &  $3.5^{+0.9}_{-0.8}$   &   $0.05^{+0.17}_{-0.04}$   &   9 \\
10  &  B1929+10   &   0.227   &    1.157   &    1.034   &  $304$  &  $28^{+5}_{-4}$ &  $4.5^{+0.3}_{-0.5}$   &   $1.26^{+0.44}_{-0.35}$   &   10 \\
11  &  J2021+4026   &   0.265   &    54.682   &    7.694   &  $281$  &  $192^{+411}_{-101}$ &  $3.6^{+0.9}_{-0.9}$   &   $0.16^{+0.57}_{-0.15}$   &   11 \\
12  &  J2043+2740   &   0.096   &    1.27   &    0.706   &  $467$  &  $358^{+...}_{-...}$ &  $1.9^{+...}_{-...}$   &   $0.01^{+...}_{-...}$   &   12 \\
13  &  B2224+65   &   0.683   &    9.661   &    5.187   &  $175$    &  $28^{+6}_{-18}$ &  $5.8^{+1.2}_{-1.2}$   &   $2.0^{+13.31}_{-0.61}$   &   13 \\

  & & & & & & & & &  \\
        \hline
  \end{tabular}
  \end{center}
  {\bf Notes.} The individual columns are as follows: (1) number of the pulsar, (2) pulsar name, (3) rotational period, (4) period derivative, (5) dipolar component of the magnetic field at the polar cap,  (6) radius of the diploar polar cap, (7) radius of the hot polar cap, (8) temperature of the polar cap, (9) magnetic field strength at the polar cap, (10) reference.\\
 \noindent  {\bf References.} (1) \citealt{PAPMSK12}, (2) \citealt{MVKZCC07}, (3) \citealt{TO05}, (4) \citealt{KPZR05}, (5) \citealt{GHMGZM08}, (6) \citealt{MTET13}, (7) \citealt{SMG15,SGZHMGMX17}, (8) \citealt{PAPMSK12}, (9) \citealt{OKMC04}, (10) \citealt{MPG08}, (11) \citealt{LHHWHTTSWCC13}, (12) \citealt{BWTJDHZ04}, (13) \citealt{HHTTTWC12}}
\label{tab:PSR_obs}
\end{table*}

\subsection{Thermo - magnetic interactions at the polar cap of radio pulsars}

\noindent There is perhaps no place at and in a neutron star, where the magnetic and thermal evolutions are more intense coupled than at the polar cap of radio pulsars.  Magneto - thermal processes at the polar cap run on much shorter timescales and within a much smaller spatial region than in case of the magneto - thermal interactions  shortly described above.
It is widely accepted that pulsars create their radio emission by charges which are accelerated to 
ultra-relativistic energies either in a space charge limited flow of electrons \citep{AS79} or in a 
vacuum gap \citep{RS75} just above the pulsar's polar cap.
Already \cite{RS75} and \cite{AS79} noted that for a sufficiently powerful creation of electron-positron 
pairs the magnetic field at the surface of the polar cap must be significantly more curved than the far 
above the neutron star dominating dipolar field. The latter has a curvature radius of $\sim 10^8$cm, 
while copious pair production requires curvature radii $\lesssim 10^6$cm. As can be argued by the observations presented in Tab.~\ref{tab:PSR_obs}, not the standard dipolar cap, defined by the last open dipolar magnetic field line, but the much smaller bombarded one  is the area where a strong and small scale field dominates. It joins in a distance of $\sim 10$ km  field lines of the long - range dipolar field. A mechanism that creates the strong and small scale field structures at the polar cap surface could be the crustal Hall drift. It may create the required poloidal field structure out of a rich reservoir of magnetic energy stored in a toroidal field located deep in the crust and in the outer core layers \citep{GGM13,GV14}.
Pair creation and subsequent acceleration by the gap electric field component parallel to the magnetic 
field causes a backflow of electrons/positrons, while the positrons/electrons escape toward the light cylinder and produce eventually the observed emission in the radio beam. The interplay between thermal and magnetic processes at the polar cap is related to the cohesive energy. It increases strongly  with the local magnetic field strength in the cap matter \citep{ML07}. In  order to allow the formation of an accelerating gap above the polar cap, the cohesive energy (controlled by the local magnetic field) has to dominate the thermionic outflow of ions or electrons (controlled by the local temperature). The polar cap is periodically, with a period of $\sim 10\mu$s, the duration of a sparking cycle,  heated up to a few million Kelvin by the heavy bombardment with backflowing charges. To enable the periodic rebuilding of the accelerating gap, thermostatic regulation takes place. Its threshold is set by the polar cap magnetic field $B_s$ \citep{SMG15,SGZHMGMX17}. A schematic presentation of a sparking cycle is shown in Fig.~\ref{fig:thermostat}.\\

\subsection{Simultaneous X-ray and radio observations}

For some radio pulsars simultaneous radio and X-ray observation are achievable. In that case the observation of thermal X-rays originating from the polar cap region, helps to understand the basic process of radio emission, the creation of an inner accelerating region above the polar cap. In Tab.~\ref{tab:xray_all} are data of 13 pulsars collected, which have been observed both in radio and X-rays. The radio observations provided the rotational period $P$, its time derivative $\dot{P}$, the dipolar surface magnetic field strength $B_d \propto \sqrt{P\dot{P}}$, and the radius of the conventional polar cap, determined by the last open field line of the dipolar magnetic field $R_{dip}\propto\sqrt{R^3/P}$, where $R$ denotes the radius of the neutron star. These ``radio-data'' are measured with high precision, so no error bars appear.

\noindent In contradiction, surface temperatures and areas of the thermal photons emitting regions, obtained by blackbody fits from X-ray spectra, have large systematic uncertainties. Therefore, in almost all cases large error bars have to be applied to the radius of the bombarded hot polar cap area $R_{pc}$ and its surface temperature $T_s$. According to the PSG  this hot area is the basis of the inner acceleration region. It is located within the conventional polar cap and the small scale $B_s$ can join the dipolar field $B_d$ at a height of a few kilometer above the surface. Then, flux conservation arguments yield for the local magnetic surface field $B_s=B_d\times (R_{dip}/R_{pc})^2$ values that are also tainted with large errors. 

\subsection{Millisecond pulsars}

However, there is still a problem that can not be completely reconciled with the PSG. It concerns the radio emission of millisecond pulsars (MSP), which are supposed to have poloidal dipolar surface field strength of $\lesssim 10^9$ G.  If for the creation of an inner accelerating region above the polar cap a local surface magnetic field strength $\gtrsim 10^{13}$ G is necessary, the argument of flux conservation does not provide such strong small scale field components. As an example may serve   J0437-4715, the nearest known MSP. From its rotational period $P\approx0.0058$ s the radius of the conventional polar cap, defined by the last open dipolar field line, is $R_{pc}^{dip}\approx 1.9\times 10^5$ cm; with $\dot{P}\approx 5.7\times 10^{-20}$ a dipolar surface magnetic field $B_s^{dip}\approx 1.2\times 10^9$ G can be estimated. Recent XMM - observations of this MSP were analyzed by \cite{B13}. The best fit of the X-ray spectrum was achieved with a combination of three blackbody plus one power law component. The radius of the hottest emission region was fitted to $R_{pc}\approx 4\times 10^3$ cm,  the surface temperature there $T_s\approx 3.2\times 10^6$ K. Flux conservation then returns a surface field at the hottest emission region $B_s\approx 2.6\times 10^{12}$ G, too small to guarantee for a sufficiently high cohesive energy in the surface layer of the polar cap. A similar result, $B_s\approx 1.2\times 10^{13}$ G can be derived from the evaluation of XMM-observations of the MSP J1614-2230 provided by \cite{PWBCGHJMR12}. Thus, perhaps a mechanism other than the Hall drift must be at work and/or is ``additionally'' active to create strong and small scale field components at the by heavy bombardment continuously heated polar cap area. The proposal is that thermoelectric interaction could indeed be this mechanism. 

\section{Thermoelectricity at the pulsar polar cap}

\begin{figure*}
\centering
\includegraphics[width=8cm, height=6cm]{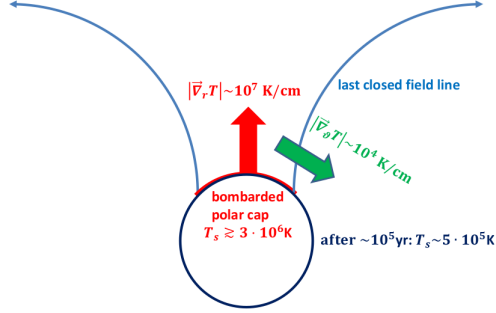}
\caption{Schematic presentation of the physical situation at the polar cap.}
\label{fig:Phys_Sit_PC}
\end{figure*}

\begin{figure*}
\centering
\includegraphics[width=5.cm, height=7cm]{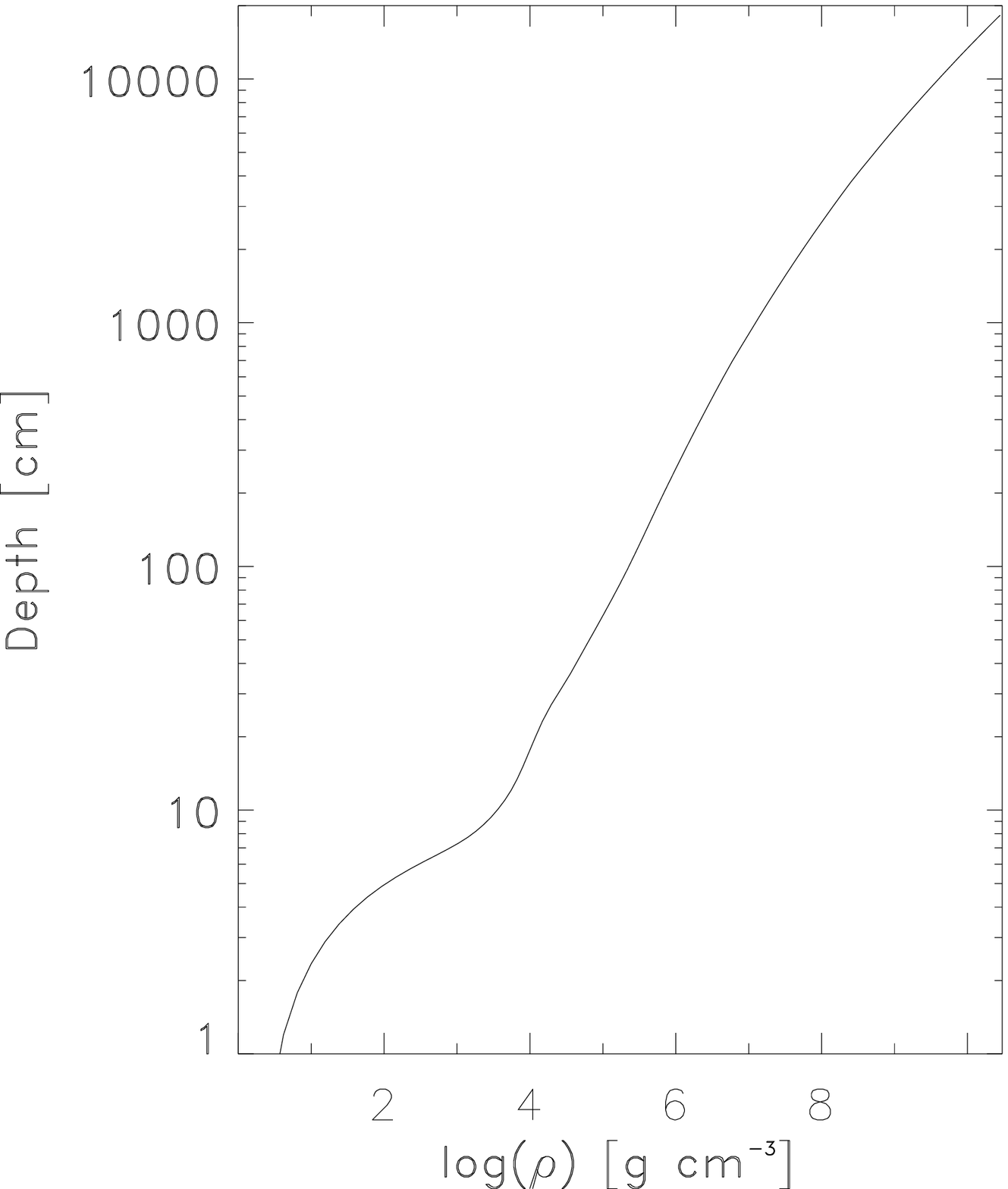}
\includegraphics[width=5.cm, height=7cm]{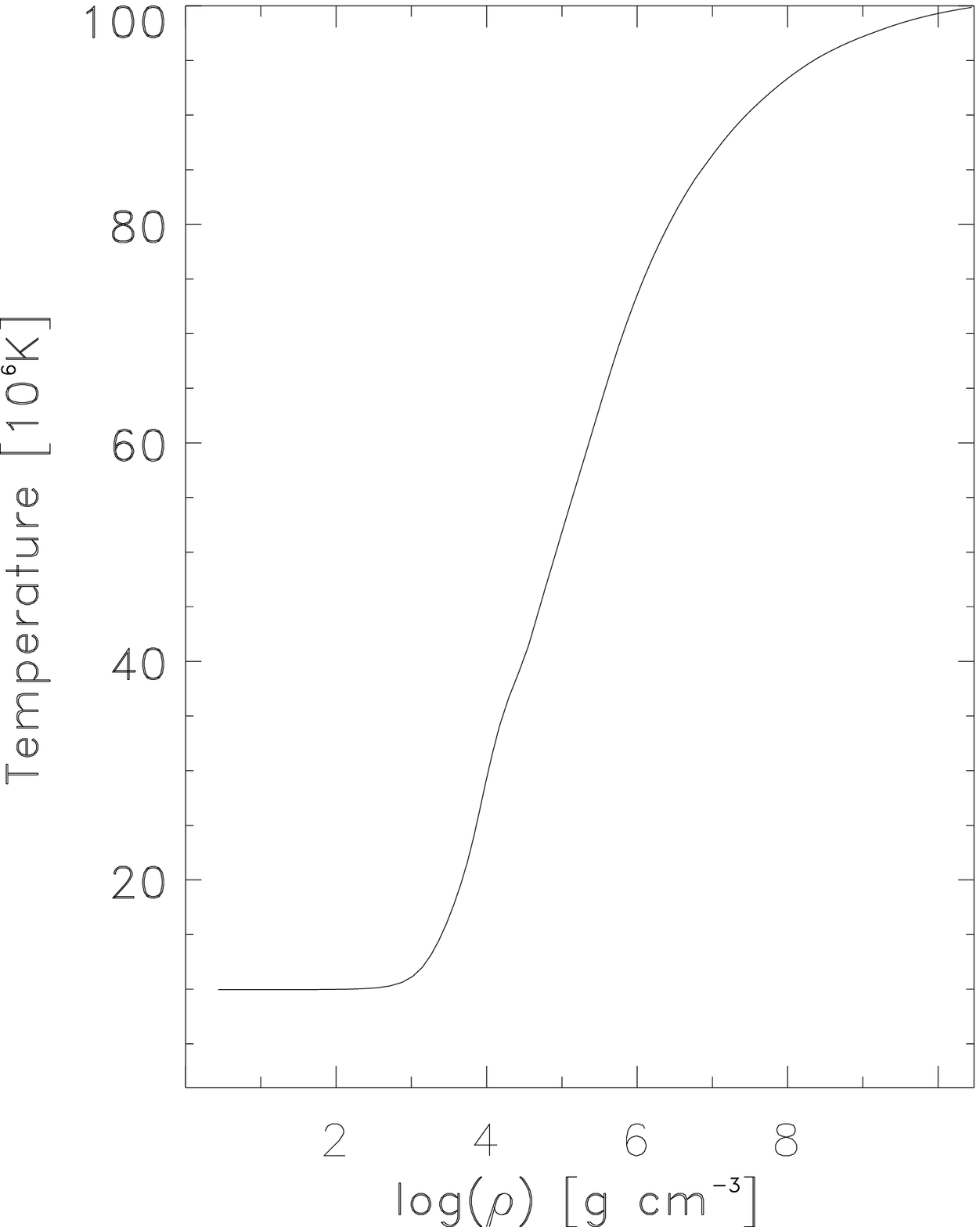}
\includegraphics[width=5.cm, height=7cm]{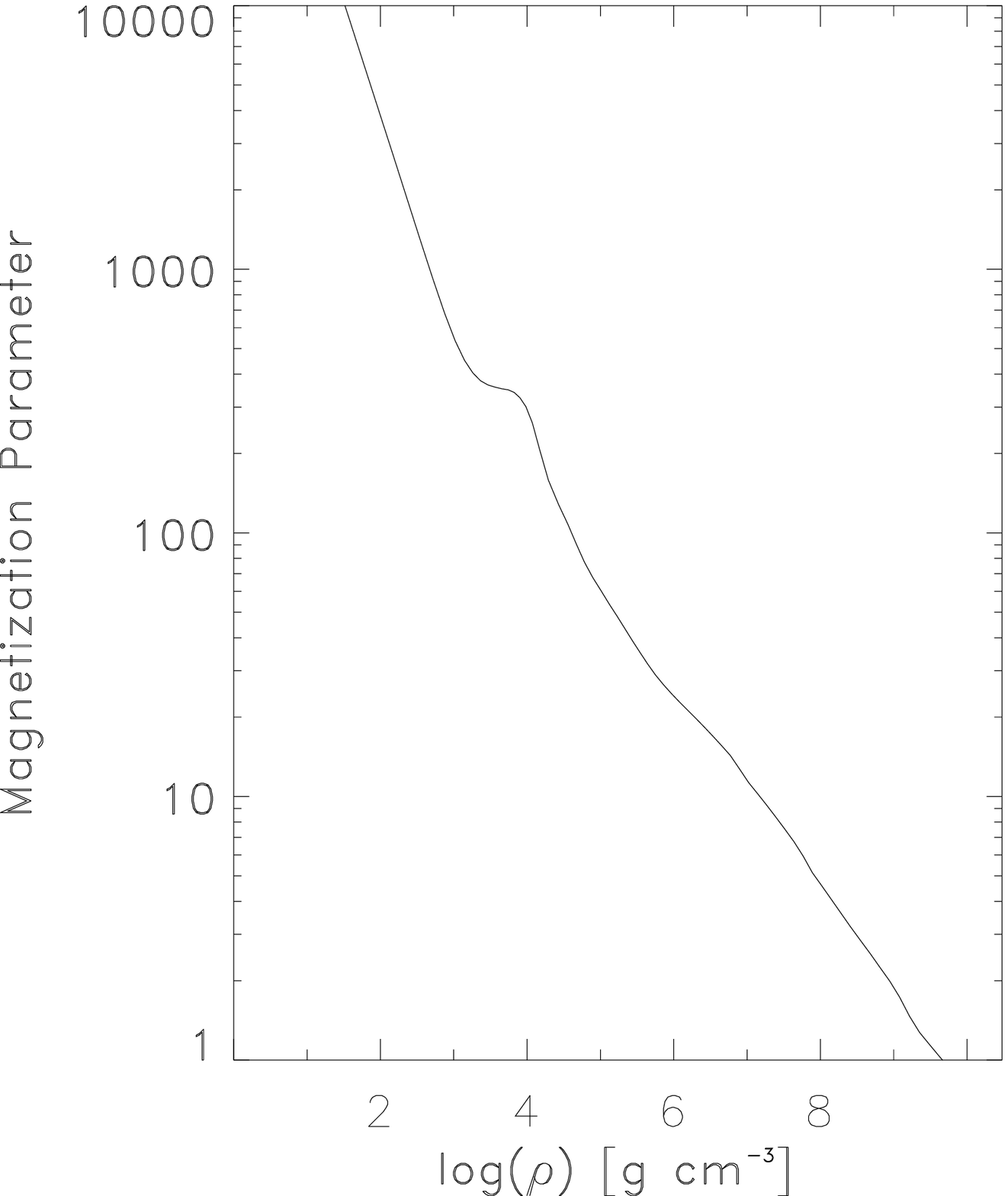}
\caption{Depth below the surface of the polar cap (left), temperature (middle), and magnetization parameter (right) as functions of the density.}
\label{fig:pol_cap_struc}
\end{figure*}

\begin{figure*}
\centering
\includegraphics[width=7.cm, height=7cm]{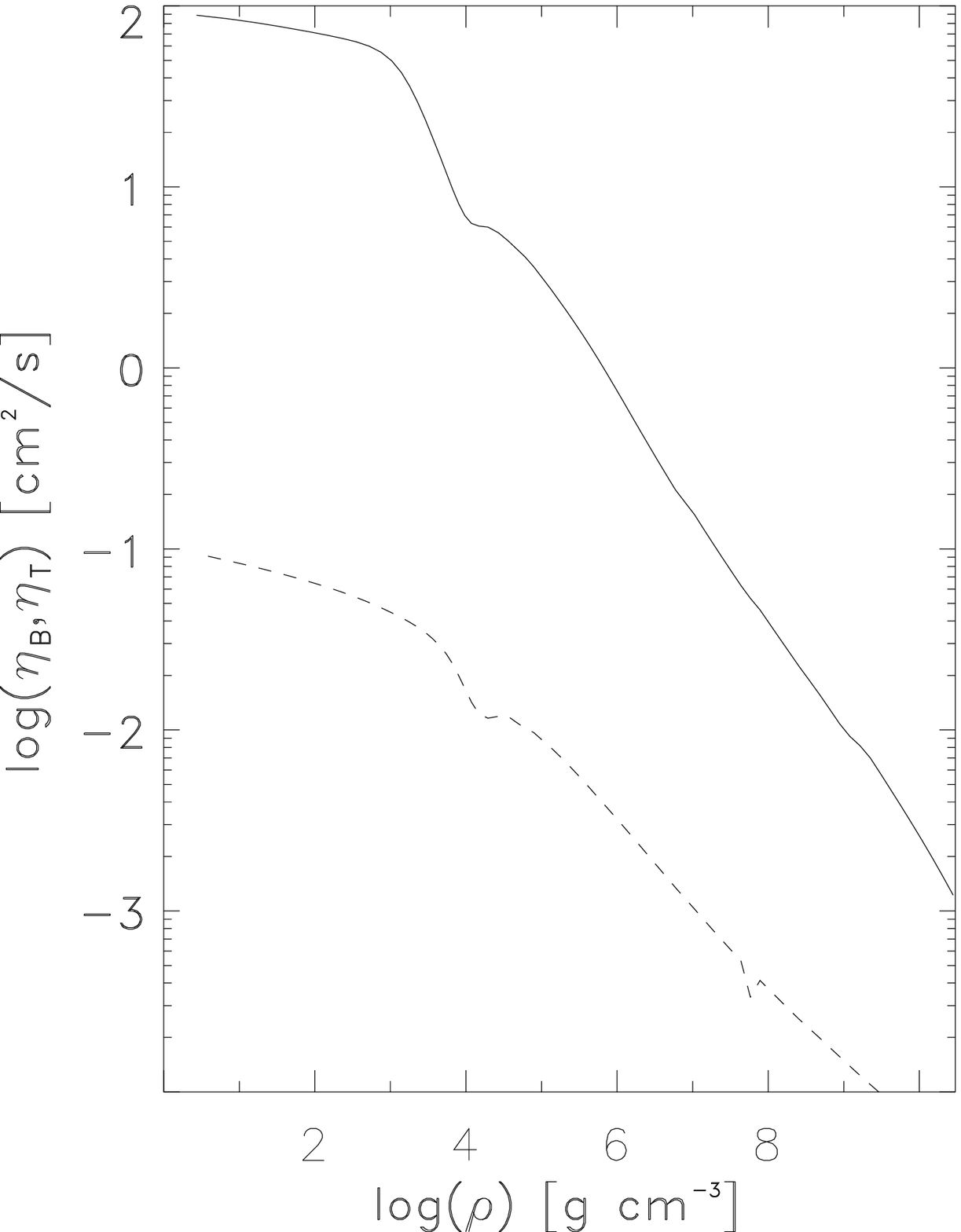}
\includegraphics[width=7.cm,height=7cm]{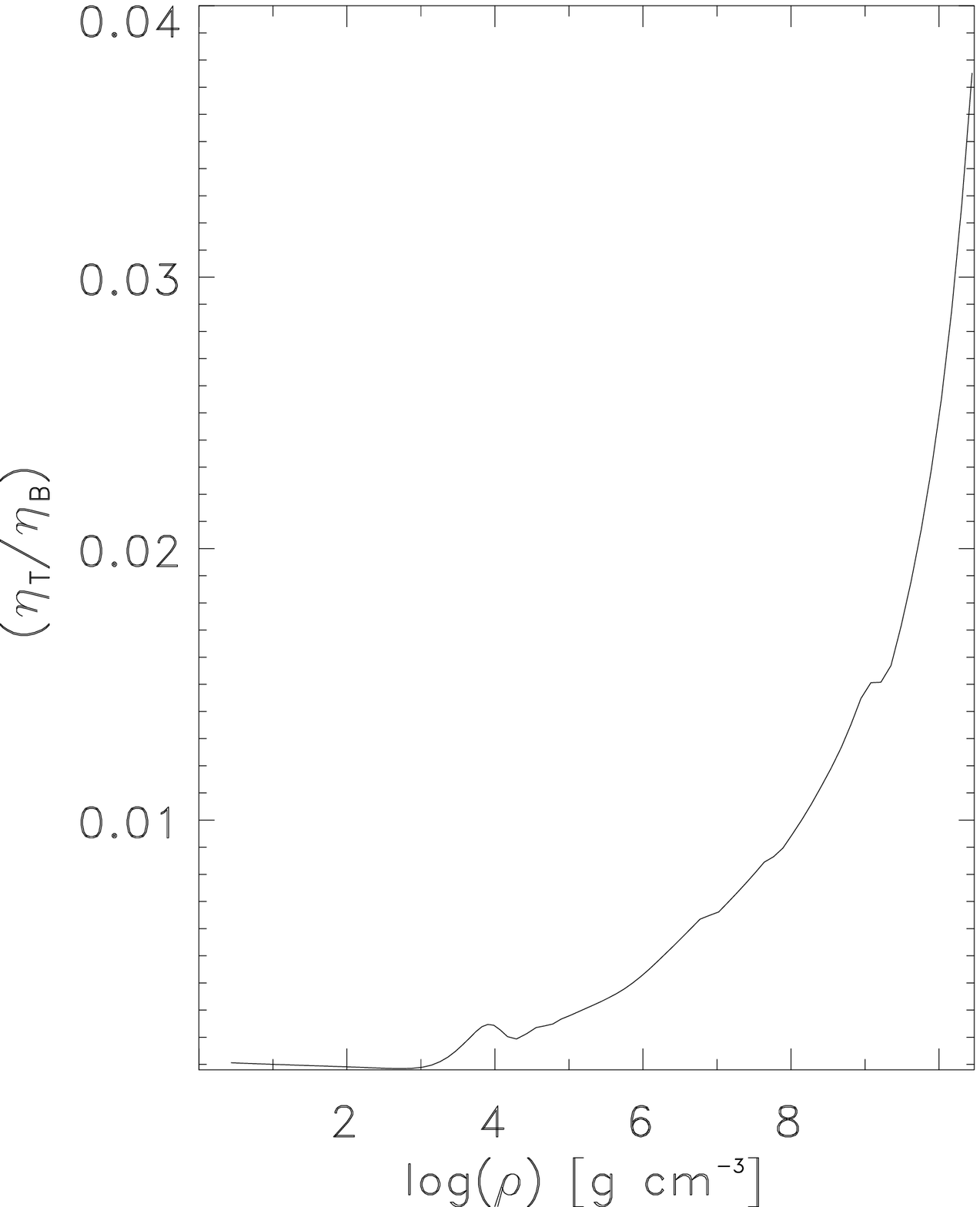}
\caption{ Thermal (dashed) and magnetic (full line) and diffusivities (left), and their ratio  as functions of the density in the envelope (right).}
\label{fig:pol_cap_phys}
\end{figure*}

An idea about the physical situation at and around a radio pulsar's polar cap is sketched in Fig.~\ref{fig:Phys_Sit_PC}. Striking features are the big temperature differences between the polar cap and the rest of the neutron star surface and to its surrounding. These great differences cause enormous temperature gradients both in radial and meridional direction, which are, however, restricted to a relatively shallow layer of the envelope beneath the surface. 
\noindent The lifetime of radio pulsars is typically $10^6\ldots 10^7$ yr. Over this lifetime, the polar cap of radio pulsars is continuously bombarded with backflowing ultrarelativistic charged particles. This  makes this small area significantly hotter than the remaining surface of the neutron star. The rest surface cools according to quite well understood cooling scenarios (URCA or DURCA, photons), thereby increasing with time the meridional temperature gradient. After $\sim 1$ Myrs the big part of the surface has a temperature in the order of a few $10^5$ K \citep{PGW06,VRPPAM13}.\\
\noindent The magnetic field penetrating the  polar cap surface is dominated by its strong 
radial components. These exert a huge thermally insulating effect at the rim of the  polar cap. An important issue is the surface density at the hot polar cap, where the neutron star matter is condensed - either liquid or solidified. A guess is the so-called zero pressure density. It increases with the local magnetic field strength  proportional to $B_{12}^{6/5}$ \citep{DL01}. For a local magnetic field of $10^{12}\ldots 10^{14}$ G  zero pressure densities  range from $\rho_s \sim 10^4\ldots 10^6$ g cm$^{-3}$. At such densities and magnetic field strengths the magnetization parameter $\omega_B\tau$ is large enough to guarantee the maintenance of a big meridional temperature gradient over the whole radio pulsar lifetime. Whether the matter at $\rho_s$ is liquid or solid decides the temperature $T(\rho_s)$ and the Coulomb coupling parameter. For sufficiently high temperatures, the surface will be in a liquid state (see \cite{PC13}).
\noindent An estimate of the radial temperature gradient relies on the fact that the heat produced by the bombardment with charged particles is released in a very shallow layer beneath the polar cap surface \citep{CR77,CR80}. It is not conducted inwardly, but radiated away almost immediately. Therefore, this temperature gradient can be estimated from $\frac{dT}{dr}=\kappa^{-1}\sigma_{SB}T_s^4$, with $\sigma_{SB}$ being the Stefan-Boltzmann constant. The heat conductivity $\kappa$ in the Iron polar cap at a density of 
$\rho \approx 10^6$ g cm$^{-3}$ and $T\sim 3\times 10^6$ K is taken from Fig. 5 of \cite{PBHY99}, not taking into account quantizing effects of the magnetic field. With $\kappa\approx 10^{14}$ erg/(cm s K) is $|\frac{\partial{T}}{\partial{r}}|\approx 10^7$ K cm$^{-1}$.  Because the magnetization parameter $\omega_B\tau$ below densities of $\rho\sim 10^5$g cm$^{-3}$ exceeds $100$, the heat flux in meridional direction is suppressed by at least a factor $10^4$ in comparison to the radial one and the corresponding temperature gradient is stably huge. Since almost all heat is released in a very shallow layer beneath the polar cap, the meridional temperature gradient is strongest close to the cap surface and decreases rapidly with increasing depth.
The meridional temperature gradient is determined by the huge temperature difference between the hot polar cap ($T_s\approx 3\times 10^6$ K and the rest surface ($\approx 5\times 10^5$ K after $10^6$ yrs of pulsar lifetime). Given the strong thermally insulating effect of the radial magnetic field at the rim of the hot polar cap, the drop of the surface temperatur will proceed along a very small meridional distance, say a few meter. Therefore, the meridional temperature gradient can be estimated as $|\frac{\partial{T}}{R\partial{\theta}}|\lesssim 10^4$ K cm$^{-1}$.
\noindent Such a large temperature gradient may clearly provide  the potential for a thermoelectric modification of the magnetic 
field structure of the polar cap. Since it is restricted onto the small, perhaps only a few centimeter thick ring along the 
rim of the cap, it will modify the field preferentially in that region.\\
\noindent The geometry of the physical situation at the polar cap is settled by a strong meridional temperature gradient, an
almost radial gradient of the thermopower coefficient, and an also almost radial magnetic field penetrating the cap surface. This suggests that by thermoelectric interactions an azimuthal magnetic field component may be preferentially amplified. The typical length 
scale of this component, its curvature radius, is the radius of the polar cap, typically $30 \dots 300$ m (see Tab.~\ref{tab:xray_all}).\\ 
\noindent In order to get an idea about the physical situation at the polar cap region a 2D magneto - thermal envelope model has been considered by use of the corresponding code of the Alicante group and the code calculating magnetized transport coefficients provided by the St. Petersburg group (see \cite{PC13,PPP15}, http://www.ioffe.ru/astro/conduct/). The density, temperature, and magnetization parameter profiles are shown in Fig.~\ref{fig:pol_cap_struc}. At the bottom of the envelope, at $\rho=10^{10}$ g cm$^{-3}$, the temperature is assumed to be about $10^8$ K.  A dipolar poloidal magnetic field of $B \sim 10^{12}$ G polar surface strength penetrates almost radially the envelope beneath the polar cap and the cap itself. 
\noindent Fig.~ \ref{fig:pol_cap_phys} depict the "initial" conditions of the thermal and magnetic diffusivities  in the polar envelope region, i.e. before via thermoelectric interactions a   $B_ {\varphi}$ at the rim of the polar cap has been created and subsequently modifications of the poloidal field at the polar cap surface may occur.

\subsection{Induction equation with thermoelectric field}
The induction equation which takes into account Ohmic diffusion, Hall drift, and thermoelectric effects
reads 

\beq 
\frac{\partial\vec{\tilde B}}{\partial t} &=&-\nabla\times\bigl\{\eta_B\left[\nabla\times\vec{\tilde B}
+\omega_B\tau\left(\nabla\times\vec{\tilde B}\times \vec{b}\right)\right] 
\nonumber \\
&-& \eta_T\bigl[\nabla\tilde{T}+\frac{\eta_{\perp} - 
\eta_T} {\eta_T}\left({\vec b}\times\left(\nabla\tilde{T} 
\times\vec b\right)\right)  
\nonumber \\ 
&+& \frac{\eta_{\wedge}}{\eta_T}\left(\nabla\tilde{T} \times \vec{b}\right) \bigr] \bigr\} \;,
\label{eq:IndEq}
\eeq

\noindent where $\vec{\tilde B}=\vec{B}/B_0$ and $\tilde T=T/T_0$ are the normalized magnetic field and 
temperature; $\vec b=\vec B/B$, the unit vector of the magnetic field. For the estimates presented here, the normalizations $B_0=10^{12}$G and $T_0=10^6$K are used. 
The magnetic diffusivity is 

\be
\eta_B=\frac{c^2}{4\pi\sigma}\;,
\label{eq:mag_diff}
\ee

\noindent with $c$ being the velocity of light and $\sigma$ denoting the scalar electric conductivity, i.e. the conductivity parallel to the magnetic field. Here, the influence of the thermal  on the magnetic evolution  proceeds only via the temperature dependence of the scalar electric conductivity. The factor in front of the Hall term, $\eta_B\omega_B\tau$, depends only on the electron density. 
The electric field proportional to the thermal diffusivity $\eta_T$, describes the thermoelectric influence on the magnetic field. A pure battery effect is given by $\eta_T\nabla\tilde{T}$. The scalar, parallel to the magnetic field component of the thermal diffusivity is 

\be
\eta_T=\frac{cQ_jT_0 T_6}{B_{12}B_0}\;.
\label{eq:therm_diff}
\ee

\noindent $Q_j$ is the scalar
thermopower, $T_6=T/10^6$K, $B_{12}=B/10^{12}$G. The components of the
thermal diffusivity tensor perpendicular to the magnetic field ($\eta_\perp$) and the off-diagonal Hall
component ($\eta_\wedge$) as well as the scalar $\eta_B$ and $\eta_T$  are calculated by use of the code of the St. Petersburg group (http://www.ioffe.ru/astro/conduct/). The components of the thermal diffusivity $\eta_\perp$ and $\eta_\wedge$ depend  via the magnetization parameter $\omega_B\tau$ on the magnetic field  (see \cite{UY80, PY96}). Thus, the temporal and spatial evolving thermoelectrically modified magnetic field will cause a time dependence of the coefficients in Eq.~\ref{eq:IndEq}; the field will evolve non-linearly.\\ 

\subsection{Rough estimates}
Considering the leading terms in Eq.~\ref{eq:IndEq}, an estimate about attainable saturation strength of the
thermoelectrically generated azimuthal field $B_ {\varphi}$ and its typical growth time can be performed.  Although
the Hall drift may dominate the Ohmic diffusion/dissipation, the competition goes mainly between thermoelectric battery 
and Ohmic decay. Starting with a purely poloidal magnetic field at the polar cap, in axial symmetry the Hall drift does not contribute to the creation of  $B_ {\varphi}$ and is not dissipative. However, it may compress the "initially" almost radial polar cap magnetic field thereby increasing its curvature and accelerating the Ohmic dissipation. Also, the thermal drift of the magnetic field, as the Hall drift a non-linear process, is not taken into account for this rough estimate. Then the necessary condition for a significant thermoelectric field generation is the dominance of the battery effect over the Ohmic decay

\be
\eta_B\nabla\times(\nabla\times B_{\varphi})+ \frac{\partial{\eta_B}}{\partial{\rho}}\frac{\partial{\rho}}{\partial{r}}\times(\nabla\times B_{\varphi})  < \frac{\partial{\eta_T}}{\partial{\rho}}\frac{\partial{\rho}}{\partial{r}}\times\mbox{grad}_\theta\tilde{T}\;,
\label{eq:cond1}
\ee

\noindent where $B_{\varphi}$ is the azimuthal magnetic field strength normalized to $B_0$ . Approximating the scale of $B_ {\varphi}$, $L_B$, by the radius of the polar cap and  the scale of  the meridional temperature gradient by $L_T$ the condition above simplifies to

\be
\eta_B\frac{B_ {\varphi}}{L_B^2}+\frac{\partial{\eta_B}}{\partial{\rho}}\frac{\partial{\rho}}{\partial{r}}\frac{B_ {\varphi}}{L_B} <  \frac{\partial{\eta_T}}{\partial{\rho}}\frac{\partial{\rho}}{\partial{r}}\frac{\tilde{T}}{L_T}\;.
\label{eq:cond2}
\ee

\noindent The diffusivities and their derivatives can be read from Figs.~\ref{fig:pol_cap_struc} and \ref{fig:pol_cap_phys}. Taking the values at $\rho = \rho_s\sim 10^4$gcm$^{-3}$, the condensation density for $B_0$, one finds $\eta_B\sim 6$ cm$^2$s$^{-1}$,  $\frac{\partial{\eta_B}}{\partial{\rho}}\frac{\partial{\rho}}{\partial{r}} \sim 10^{-1}$cm s$^{-1}$ and $\frac{\partial{\eta_T}}{\partial{\rho}}\frac{\partial{\rho}}{\partial{r}} \sim 10^{-3}$cm s$^{-1}$. With this and a typical $L_B\sim 10^4$cm (see Tab.~\ref{tab:PSR_obs}) it turns out that the first term in Eqs.~\ref{eq:cond1} and \ref{eq:cond2} can safely be neglected in comparison to the second one. Therefore, the condition that thermoelectric field generation is not dominated by Ohmic decay reads as

\be
\left(\frac{\frac{\partial{\eta_T}}{\partial{\rho}}}{\frac{\partial{\eta_B}}{\partial{\rho}}}\right)\frac{L_B}{L_T}\frac{\tilde{T}}{B_ {\varphi}} >1\;.
\label{eq:growth_cond}
\ee

\noindent Given the estimates of the partial derivatives of the diffusivities, $\tilde{T}\approx 3$ (see Tab.~\ref{tab:PSR_obs}), and $B_ {\varphi}\sim 1$, Eq.~\ref{eq:growth_cond} requires that the scale length of the meridional temperature gradient must be $L_T\sim 10^{-2}L_B$, i.e. at most a few meter.  Although the thermally insulating effect of the poloidal magnetic field at the rim of the polar cap is certainly very strong and the evolving azimuthal field will enforce this insulation additionally, it remains an open question as to whether such a strong meridional temperature gradient indeed can be maintained over the lifetime of radio pulsars. Clearly, this estimate has to be considered with great caution. Especially the observed radii of the hot polar cap vary between $\sim 10$ and $\sim 300$ m and are themselves afflicted with large systematic errors of the blackbody fits (see Tab.~\ref{tab:PSR_obs}). Therefore, the estimation of a potentially achievable saturation strength of the azimuthal field along the rim of the polar cap on the order of $10^{12}$G serves only as an encouragement to carry out a more detailed study which also takes into account the Hall drift and the thermal drift of the magnetic field.

\noindent An estimate of the growth time of the thermoelectrically generated azimuthal field $B_ {\varphi}\sim 1$ provides the battery term:

\be
\frac{\partial B_ {\varphi}}{\partial t} \sim  {\mbox{grad}_r}\eta_T   \times {\mbox{grad}}_{\theta}\tilde T\;.
\label{Eq:growth_time}
\ee

\noindent With  ${\mbox{grad}}_r\eta_T=\frac{\partial{\eta_T}}{\partial{\rho}}\frac{\partial{\rho}}{\partial{r}} \sim 10^{-3}$ cm s$^{-1}$ and ${\mbox{grad}_\theta}\tilde{T} \sim 3\cdot 10^{-2}$ cm$^{-1}$ a typical growth time $\tau_{B_{\varphi}} \sim 3\cdot 10^4$ s can be found, while the corresponding Ohmic decay time is only by a factor $\approx 3$ larger. Therefore, this assessment does also not provide a clear decision as to whether the thermoelectric field creation dominates the Ohmic decay.  In any case  $\tau_{B_{\varphi}}$ is significantly shorter than the typical lifetime of radio pulsars.\\
\noindent The main interaction of the thermoelectrically created $B_ {\varphi}$ with the poloidal field which penetrates the polar cap surface will certainly proceed via the Hall drift. Similar to the effect of dipolar toroidal field components with maximum field strength in the equatorial plane \citep{VRPPAM13,GV14}, the toroidal field around the rim of the bombarded polar cap will shift the poloidal field components at the polar cap surface toward the center of the cap. During this process, the poloidal components will be compressed  around the magnetic pole, which may result in an increase both of curvature and field strength of $B_s$.  This is, however, a crude and speculative guess. A careful numeric simulation of the Hall drift driven modification of the polar cap field towards the for radio emission required field structures is necessary.

\section{Conclusion}

The richness in species of magneto - thermal interactions and the short description of their basic nature should demonstrate  that the combinations of the magnetic and thermal evolution in neutron stars are very diverse. The physics of these interactions is subject of an intense scientific activity which will certainly continue in the future. There are serious open problems that are connected with an extension to three dimensional studies of the coupled magnetic and thermal evolution. Another open issue is the coupling of these evolutions to hydrodynamics, when either processes in the upper and hot layers of  young neutron star envelopes or field re-organization processes in the liquid core are concerned. The latter is additionally complicated by the appearance of superfluidity and superconductivity.\\
\noindent One of these unsolved problems is the potential  occurrence of thermoelectric interactions at the polar cap of radio pulsars. The here presented rough estimates  suggest that thermoelectric interactions may play a role in the evolution of the surface magnetic field structures at the polar cap of radio pulsars. It seems possible that the meridional temperature gradient is large enough and the geometry of the physical situation at the polar cap is so favorable that on relatively short time scales a strong toroidal field along the rim of the cap can be generated. A conceivable scenario is that thermoelectric field amplification ``collaborates'' with the crustal Hall drift to affect the poloidal field structure, making it so strong and curved that the conditions for ample electron-positron pair creation are met.\\
\noindent However, there are many and serious caveats in this scenario of polar cap field evolution. The most striking one is the correct calculation of the transport coefficients in the magnetized matter at densities $\rho < 10^{10}$ g cm$^{-3}$.  In such low density regions the magnetic field has a quantizing effect which leads to oscillations in the transport coefficients. At densities  $\rho < 10^{6}$ g cm$^{-3}$ the matter is not completely ionized and the electron gas is only partially degenerated.  Since the state of the envelope matter depends on the local magnetic field strength, both the transition from liquid to solid matter and the value of the zero pressure density will depend on the state of magnetic field evolution at the polar cap. The Coulomb coupling parameter which determines the state of aggregation of the envelope matter depends on the magnetic field strength. This means that liquid and solid regions can vary their location as the field strength varies \citep{PC13}. Given the high surface temperature of $T_s\sim 3\times 10^6$ K, it is quite possible that parts of the polar cap surface are in a liquid state. This complicates the modeling of the magnetic field evolution there drastically, because raising Lorentz forces can drive circulations in the liquid. A realistic modeling requires to study the magneto - thermal evolution together with hydrodynamical motions at the polar cap. A code that solves the coupled differential equations of induction, thermal transfer and hydrodynamics in partially ionized and partially degenerated matter is not yet developed.\\
\noindent Because the rough estimates presented above do not clearly demonstrate that thermoelectric field generation indeed will play an important role for the creation of the required polar cap magnetic field structure, in a next step the influence of Hall and thermal drift onto the magnetic field evolution will be studied. This should return an information how strong the meridional temperature gradient must be in order to drive  thermoelectric field growth.  The here presented sketch of a scenario can only serve to justify and to encourage further, much more detailed studies.\\

\section*{Acknowledgments}
I am grateful to J.A.Pons, who provided the thermal and magnetic diffusivities by use of the St. Petersburg code and the 2D envelope simulation of the situation at the polar cap. Gratefully acknowledged are numerous discussions with J.A. Pons and A. Potekhin, who brought to my attention the caveats of the here presented idea. Enlightening discussions with A. Szary and G. Melikidze are gratefully acknowledged too. 

\footnote{}

\bibliography{pulsars}
\end{document}